\documentclass[acmsmall,screen]{acmart}

\usepackage{graphicx}
\usepackage{hyperref}
\usepackage{amsmath}
\usepackage{commath}
\usepackage{algorithm}
\usepackage[noend]{algpseudocode}
\usepackage[tikz]{bclogo}
\usepackage{tikz}
\usetikzlibrary{calc}
\usepackage{pgfplots}
\pgfplotsset{compat=1.17}
\usetikzlibrary{positioning}
\usepackage{tikzscale}
\usepackage{float}
\usepackage{subcaption}
\usepackage{multirow}
\usepackage{nicefrac}
\usepackage[frozencache,cachedir=minted-cache]{minted}
\usepackage{enumitem}
\usepackage{wrapfig}
\usepackage{pgf-pie}
\usepackage{float}
\usepackage{subcaption}
\usepackage{csvsimple}

\usepackage[compact,noindentafter]{titlesec}
\titlespacing{\section}{2pt}{2ex}{1ex}
\titlespacing{\subsection}{0pt}{1.5ex}{0.75ex}
\titlespacing{\subsubsection}{0pt}{1ex}{0.5ex}

\definecolor{edgeblue}{RGB}{0, 0, 200}
\definecolor{edgegreen}{RGB}{0, 200, 0}
\definecolor{gptgreen}{RGB}{0, 166, 126}
\definecolor{soorange}{RGB}{231, 133, 35}
\definecolor{scholarpurple}{RGB}{169, 1, 251}
\definecolor{bgcode}{rgb}{0.95,0.95,0.95}

\usepackage{xcolor}

\definecolor{kscolor}{rgb}{0.9,0.1,0.1}
\definecolor{mscolor}{rgb}{0.1,0.1,0.9}
\definecolor{stcolor}{rgb}{0.2,0.7,0.1}

\theoremstyle{definition}

\theoremstyle{definition}

\newcommand{\code}[1]{\texttt{#1}}
\newcommand{\str}[1]{\ensuremath{\mathtt{#1}}}
\newcommand{\Paragraph}[1]{\smallskip\noindent{\bf #1}}

\DeclareMathOperator*{\argmax}{arg\,max}

\newcommand{\toolX}{\textsc{CodeScholar}}
\newcommand{\sast}{\texttt{SAST}}

\newminted[codeExampleNoHighlight]{python}{fontsize=\footnotesize, xleftmargin=\parindent, bgcolor=bgcode, escapeinside=||, mathescape=true}

\newminted[codeExample]{python}{fontsize=\footnotesize, xleftmargin=\parindent, bgcolor=bgcode, escapeinside=||, mathescape=true, highlightlines={1}, highlightcolor=scholarpurple!15}

\newminted[gptExample]{python}{fontsize=\footnotesize, xleftmargin=\parindent, bgcolor=bgcode, escapeinside=||, mathescape=true, highlightlines={1}, highlightcolor=gptgreen!30}

\newminted[soExample]{python}{fontsize=\footnotesize, xleftmargin=\parindent, bgcolor=bgcode, escapeinside=||, mathescape=true, highlightlines={1}, highlightcolor=soorange!30}

\AtBeginDocument{%
  \providecommand\BibTeX{{%
    \normalfont B\kern-0.5em{\scshape i\kern-0.25em b}\kern-0.8em\TeX}}}

\begin{document}
\sloppy
\title{CodeScholar: Growing Idiomatic Code Examples}

\author{Manish Shetty}
\affiliation{%
  \institution{Universtiy of California, Berkeley}
  \city{Berkeley}
  \country{USA}
}
\email{manishs@berkeley.edu}

\author{Koushik Sen}
\affiliation{%
  \institution{Universtiy of California, Berkeley}
  \city{Berkeley}
  \country{USA}
}

\author{Ion Stoica}
\affiliation{%
  \institution{Universtiy of California, Berkeley}
  \city{Berkeley}
  \country{USA}
}
\renewcommand{\shortauthors}{Trovato and Tobin, et al.}
\begin{abstract}
Programmers often search for usage examples for API methods.
A tool that could generate realistic, idiomatic, and contextual 
usage examples for one or more APIs would be immensely beneficial to developers.
Such a tool would relieve the need for a deep understanding of the API landscape, augment existing documentation, and help discover interactions among APIs.
We present \toolX{}, a tool that generates idiomatic code examples demonstrating the common usage of API methods. It includes a novel neural-guided search technique over graphs that grows the query APIs into idiomatic code examples.
Our user study demonstrates that in $\approx$70\% of cases, developers prefer \toolX{} generated examples over state-of-the-art large language models (LLM) like \texttt{GPT3.5}.
We quantitatively evaluate $60$ single and $25$ multi-API queries from $6$ popular Python libraries and show that across-the-board \toolX{} generates more realistic, diverse, and concise examples.
In addition, we show that \toolX{} not only helps developers but also LLM-powered programming assistants 
generate correct code in a program synthesis setting.
\end{abstract}



\maketitle

\section{INTRODUCTION}

Suppose a programmer is learning to build an image classifier.
The programmer is aware of the \texttt{torch} library in Python and some relevant 
Application Programming Interface (API) methods like \code{nn.Conv2D} and \code{nn.ReLU},
but is unsure how to use them.
The programmer looks up the official documentation of these APIs and finds a vast list of parameters and a few usage examples:


\begin{minipage}{0.475\textwidth}
\begin{codeExampleNoHighlight}
>>> m = nn.Conv2d(16, 33, 3, stride=2)
>>> input = torch.randn(20, 16, 50, 100)
>>> output = m(input)
\end{codeExampleNoHighlight}
\end{minipage}%
\hspace{0.015\textwidth} 
\begin{minipage}{0.475\textwidth}
\begin{codeExampleNoHighlight}
>>> m = nn.ReLU()
>>> input = torch.randn(2)
>>> output = m(input)
\end{codeExampleNoHighlight}
\end{minipage}

However, these examples don't describe the \textit{realistic} and \textit{common} usage of these APIs.
Specifically, the programmer wants to know the real-world usage of these API methods.
Ideally, a tool that returns a few code snippets (shown below) demonstrating
some ways in which these APIs are used together frequently would be highly beneficial. We call them \textit{idiomatic code examples}.

\begin{codeExampleNoHighlight}
self.feat = nn.Sequential(
    nn.Conv2d(27, 16, kernel_size=3), nn.MaxPool2d(2), nn.ReLU(),
    nn.Conv2d(16, 32, kernel_size=3), nn.MaxPool2d(2), nn.ReLU())

self.feat = nn.Sequential(nn.Conv2d(64, 4, 3, padding=1), nn.ReLU())
\end{codeExampleNoHighlight}

Research in industry, such as at Facebook~\cite{aroma, egpaper}, has found that developers 
frequently query for such API usage examples in their internal code-to-code search tools, 
although they were deployed to get recommendations to modify or improve code.
The widespread adoption of APIs has benefited developers. 
But to use them effectively, one requires a deep understanding of the API or 
access to high-quality documentation, which helps understand the API's behavior.
While the former is entirely infeasible, the latter is often limited in practice.
In addition, \cite{multiapiusage} find that these problems are exacerbated when developers use multiple APIs.
They find that while API methods are consistently used along with other API methods, their co-usage relationships
are seldom documented. 
\cite{predictablecode} identify that developers prefer predictable and natural expressions in code.
Developers use the term ``\textit{idiomatic}'' to refer to meaningful code that other developers find natural~\cite{miningidioms, hindle2016naturalness}.
We believe these properties also apply to API usage examples; they are preferable when realistic and idiomatic.
Overall, this indicates a need for \textit{developer tools that augment existing documentation with searchable idiomatic usage examples.}


\subsection{Existing Tools}\label{sec:existing-tools}

A few existing techniques could potentially be used to search for idiomatic usage examples.
First, are searchable online forums and websites like \texttt{StackOverflow}\footnote{https://stackoverflow.com/}, \texttt{ProgramCreek}\footnote{https://www.programcreek.com/} providing code examples upvoted by other users.
However, these tools are limited by popular queries from other users and also return an extensive list 
of results with no guarantees of idiomatic examples in them.
%
Recently, automated techniques have been proposed to provide programmers with good API usage examples through code search~\cite{examplore, katirtzis2018summarizing}.
However, they (1) struggle to rank the myriad of search results~\cite{workcodesearch} and (2) lack idiomaticity that emphasizes frequent over niche usage. 
E.G.~\cite{egpaper} attempts to solve this by mining multiple common usage examples. However, it focuses on single API queries and hence falls short in retrieving examples involving API co-usage. It also uses simple heuristics like minimum usage frequency that can be restrictive to newer APIs and more expressive usage examples.
%
On another front, Large Language Models (LLM) like GPT-X~\cite{gpt4, gpt3}, Codex~\cite{codex}, GitHub Copilot~\footnote{https://github.com/features/copilot/}, and CodeLLaMa~\cite{codellama} have demonstrated ability to generate code from natural language prompts~\cite{codellmmonitors, llmproptest, hotgpt}.
However, code-related tasks are quite challenging for LLMs~\cite{gpt4,llama2,bubeck2023sparks}.
A particularly concerning failure mode has been termed \textit{hallucination}, where the model generates incorrect or non-existent code.
\cite{gorilla} find that hallucination is especially true with tasks involving APIs due to the vast space of arguments and other complexities.

\subsection{CodeScholar}

\begin{figure*}[t]
    \centering
    \includegraphics[scale=0.67]{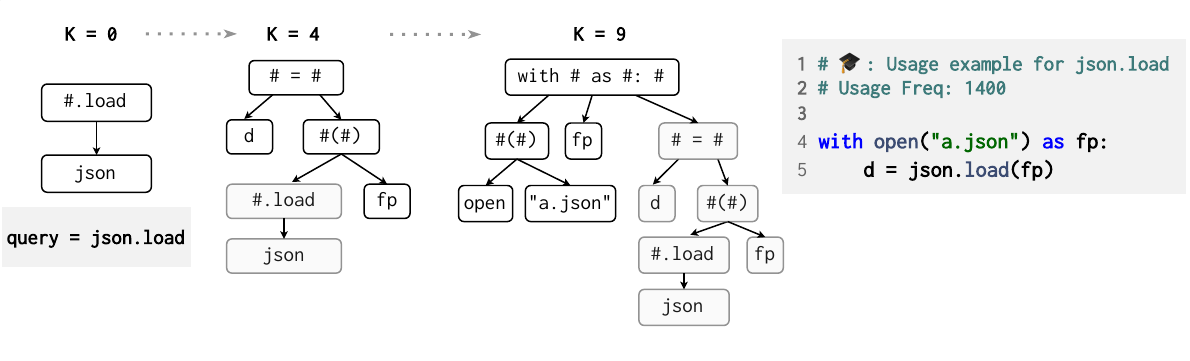}
    \vspace{-5mm}
    \caption{Overview of \toolX{}: The user's query is transformed into a simplified AST (\sast{}). Then, an iterative neural-guided search algorithm (K = iteration) adds nodes to the \sast{} to maximize the ``idiomaticity'' of the program. An in-order traversal of the resulting \sast{} retrieves an idiomatic code example.}
    \vspace{-3.5mm}
    \label{fig:overview}
\end{figure*}

This paper introduces \textbf{\toolX{}}, a novel solution to the aforementioned challenges.
\toolX{} takes a set of query API methods as input, which it then grows into full-blown code examples
using a neural-guided search algorithm.
The code examples generated are such that they contain all API methods in the query and are \textit{idiomatic}, 
i.e., non-trivial and frequently occurring in a large corpus of real-world programs.
Figure~\ref{fig:overview} illustrates this for a query API \code{\small{json.load}}.

\toolX{} capitalizes on a novel insight that idiomatic code examples are fundamentally a set of frequent subgraphs in a large corpus of programs represented as graphs.
At a high level, \toolX{} builds on a new graph abstraction of programs called Simplified Abstract-Syntax Trees (\texttt{SAST}).
Given a set of query API methods, the search starts from a set of candidate programs containing these APIs, translated to \texttt{SAST}s.
\toolX{} iteratively adds nodes to these \texttt{SAST}s while optimizing ideal properties of code examples, such as reusability, diversity, and expressivity.
Also, \toolX{} monitors these properties and stops the search at a unique convergence condition, 
allowing for a dynamic approach free from heuristics like minimum frequency or maximum program size constraints.
\toolX{} then maps \texttt{SAST}s back to programs and returns idiomatic code examples along with their usage frequency (number of times it was found in the corpus).
Below, we describe some advantages of \toolX{} through a motivating example
comparing \toolX{} to some existing solutions.

\subsubsection{\textbf{A Motivating Example.}}
\label{sec:overview}

Suppose Alice is a novice Python developer who needs a flat list 
of files from a directory with nested sub-directories. She is aware of a \code{listdir} 
function in the \code{os} library and wants to learn how to use it.
Alice searches and finds a top-rated \texttt{StackOverflow} post:
\vspace{-\abovedisplayskip}
\begin{soExample}
                                        |\textbf{StackOverflow}|
import os
path = os.getcwd()
dir_list = os.listdir(path)
\end{soExample}

While correct, she cannot use this on her directory. She then prompts \texttt{ChatGPT} \cite{gpt4}, a large-language model, for some usage examples which returns the following $3$ examples:
\vspace{-\abovedisplayskip}
\begin{gptExample}
                                            |\textbf{GPT}|
file_list = os.listdir('.')
file_list = os.listdir('/path/to/directory')
file_list = [file for file in os.listdir('.') if file.endswith('.txt')]
\end{gptExample}

These snippets are more helpful; the second one also runs on her directory, but she notices that some files are missing from the output. She's also unsure if this code is used in practice. She then turns to \toolX{} to find a more idiomatic code example. \toolX{} returns a variety of examples and their usage frequency, one of which is with frequency $130$:

\vspace{-\abovedisplayskip}
\begin{codeExample}
                                        |\textbf{CodeScholar}|
def getListOfFiles(dirName):
    listOfFile = os.listdir(dirName)
    allFiles = list()
    for entry in listOfFile:
        fullPath = os.path.join(dirName, entry)
        if os.path.isdir(fullPath):
            allFiles = allFiles + getListOfFiles(fullPath)
        else:
            allFiles.append(fullPath)
    return allFiles
\end{codeExample}

The \toolX{} generated example is more \textit{idiomatic} and solves Alice's problem. 
It not only shows what \code{os.listdir} does, but also how it fails by showing an explicit handling of nested directories.
\toolX{} generated examples are by construction grounded in real-world code, in stark contrast to language models that hallucinate~\cite{gorilla}. Consequently, the examples generated are realistic and contain context, improving the understanding of the API.
Unlike prior search-based solutions to mining idiomatic code~\cite{miningidioms, egpaper}, 
\toolX{} is language-agnostic, does not require offline mining, and supports multi-API queries.
Overall, \toolX{} usage examples are designed to be more realistic, expressive, comprehensive, and meaningful--properties of ideal usage examples.


We have implemented \toolX{} in Python and shared it as an open-source tool~\footnote{\href{https://github.com/tart-proj/codescholar}{https://github.com/tart-proj/codescholar}}.
Since providing users with good quality API usage examples is the ultimate goal of \toolX{}, we will show in Section~\ref{sec:user-study} that developers strongly prefer \toolX{} generated examples over those from state-of-the-art language models like \texttt{GPT3.5}. 
In Section~\ref{sec:eval-search}, we identify an information-theoretic inspired metric to quantify how closely generated code examples represent real-world usage. With that, we will show that for several single and multi-API queries, \toolX{} generates examples that are \emph{realistic}, \emph{diverse}, and \emph{concise}.
Lastly, in Section~\ref{sec:eval-rag} we show that \toolX{} not only helps developers, but also
AI programming assistants in a retrieval-augmented program synthesis setting.

In summary, this paper makes the following contributions:
\begin{itemize}[itemsep=0.25em]
    \item It reformulates idiomatic code examples as frequent subgraphs in a large corpus of real-world programs represented as graphs.

    \item It presents a novel neural-guided search algorithm for scalable mining of representative and idiomatic code examples for query APIs.
    
    \item It proposes a unique reusability-expressivity-diversity convergence condition for early termination of idiomatic code search.

    \item It identifies a simple information-theoretic distance metric to quantify how close generated code examples are to real-world usage.
    
    \item It realizes the above techniques in \toolX{}, a tool that supports single and multi-API queries and returns multiple usage examples with idiomaticity and provenance information.
\end{itemize}

\section{APPROACH}\label{sec:approach}

\toolX{} finds idiomatic code examples by dynamically growing graphs representing a query of API methods, as shown in Figure~\ref{fig:overview}.
\toolX{}'s new graph abstraction of programs called \sast{} is concise, concrete and maps directly to source code (Section~\ref{sec:sast}).
With this graph abstraction, \toolX{} capitalizes on a key insight that reduces idiomatic code examples to frequent subgraphs (Section~\ref{sec:problem-stmt}).
Rather than mining frequent patterns or generating code given a prompt, \toolX{} uses
a search algorithm to grow query graphs into diverse and contextually rich code examples while optimizing several ideal properties (Sections~\ref{sec:neuromatch}-\ref{sec:neural-search}).

\subsection{\sast{}: Simplified Abstract-Syntax Tree}
\label{sec:sast}

\toolX{} works on a graph representation of a program. A program's abstract syntax tree (AST) can be represented as a graph that \toolX{} could use. However, AST can have many abstract nodes, which could result in poor learning. Aroma \cite{aroma} and \cite{egpaper} use simplified parse trees to address the complexity of an AST. Simplified parse trees, created from parse trees, do not have information about node types. Moreover, simplified parse trees are created using a third-party library, which is not as robust and maintainable as Python's own \texttt{ast}  module. We propose to use simplified ASTs (called \sast{}), which associate node type information with each node in a simplified parse tree.

We formally define a Simplified Abstract-Syntax Tree (i.e., \sast{}) of a program as
a data structure to represent programs. It is recursively defined as a non-empty list whose elements could be:
\begin{itemize}
    \item a token representing keywords (such as \code{while} and \code{if}) and symbols (such as \code{+}, \code{*}, \code{:}, \code{\}}), 
    \item a token representing non-keywords (such as variable, field, and method names) or
    \item a simplified abstract-syntax tree.
\end{itemize}

\begin{figure}
    \centering
    \includegraphics[scale=0.65]{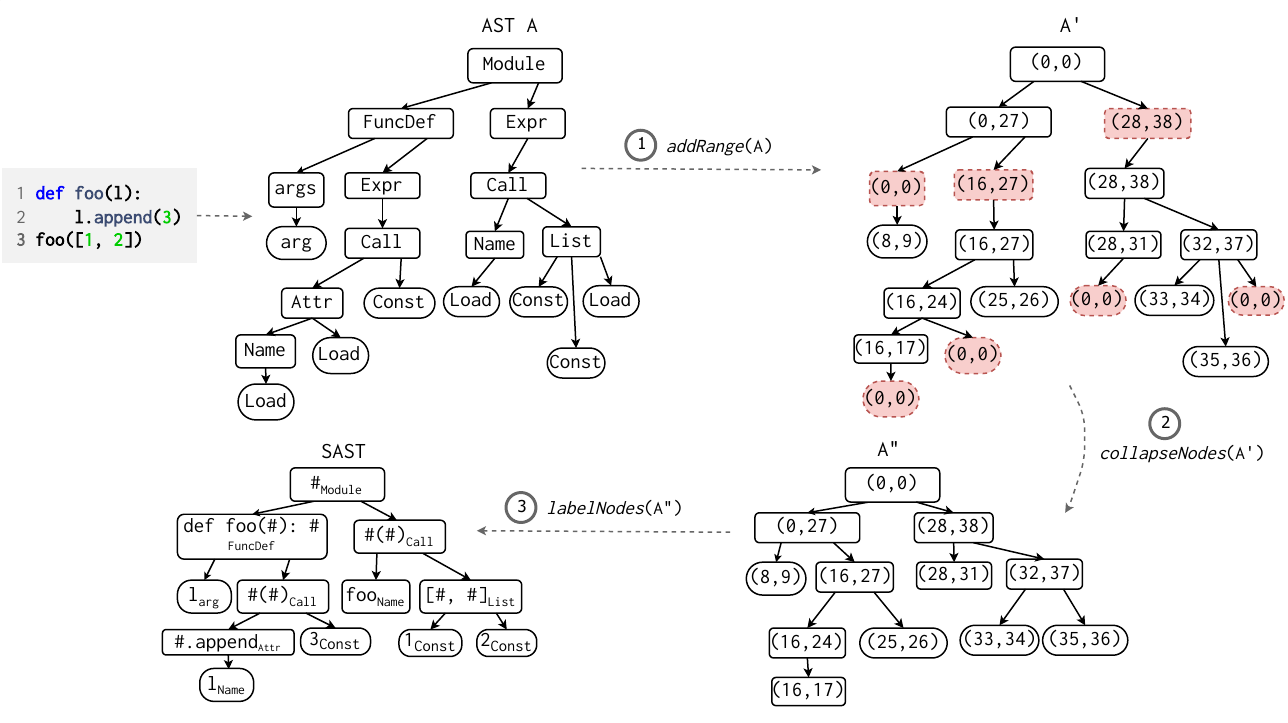}
    \caption{Transformation of an AST to a \sast{}. First, the $addRange$ transformation adds start and end code locations for each node of the AST. Then ${\rm\it collapseNodes}$ simplifies the tree by removing redundant nodes (marked red). Lastly, $labelNodes$ augments nodes with the substrings mapped to the subtree rooted at that node, resulting in the \sast{}. The \sast{} nodes retain the original AST node types along with their new labels.} 
    \label{fig:sast}
    \vspace{-3mm}
\end{figure}

Figure~\ref{fig:sast} shows an example \sast{}. Each node in the \sast{} has an associated label representing the subtree rooted at that node.
Once \toolX{} creates the \sast{} for a code snippet, the subsequent steps in \toolX{} are language-agnostic.
We now describe how \toolX{} transforms ASTs into a \sast{}s. In the following, we assume the AST to be a graph $A=(V, E)$ where $V$ represents a set of nodes and $E$ is a set of edges. We use $n \in A$ to denote a node in the set of nodes $V$ of an AST $A=(V, E)$.

First, \toolX{} adds concrete syntax information to the nodes in the AST using the function \textit{addRange}. Specifically, \toolX{} adds the start and end locations of the span of code that maps to each node, which \toolX{} calculates using line number and column offset information available in the parser-generated AST. Formally, ${\rm\it addRange}(A)$ transforms AST $A$ of program $p$ by augmenting each node $n \in A$ with a tuple $range(n) = (l, r)$, i.e., the start and end locations of the span of code corresponding to $n$ in $p$.

Next, \toolX{} collapses the AST by removing redundant nodes using the function \textit{collapseNodes}.  Specifically,  \textit{collapseNodes} collapses nodes that are too abstract (e.g., \str{withitem}, \str{arg}, etc.) to represent any relevant corresponding concrete syntax into their parent nodes.
Similarly, nodes pointing to redundant concrete syntax (relative to the parent node) are also collapsed, resulting in a simplified graph structure. \textit{collapseNodes} is implemented as follows. Let $A$ be the AST of a program $p$ transformed by $addRange$. Then, $\text{collapseNodes}(A)$ is a transformation on $A$ that performs the following modification to each node $n \in A$ with a parent node $n'$. Let $C_n$ and $C_{n'}$ be the set child of nodes of $n$ and $n'$, respectively. If $range(n) == (0,0)$ (i.e., no corresponding code span), or $n$ is a \text{node} with a single child $c$ and $range(c) = range(n)$, then $C_{n'} = C_{n'} \cup C_n$ and remove $n$ from $A$.

Finally, \toolX{} uses the function \textit{labelNodes} to augment each node in the collapsed AST with a \textit{label} representing a substring of the program mapped to the subtree rooted at that node.  The formal definition follows. Formally, for each node $n \in A$ with children $C$, the \textit{label} for $n$ is obtained by taking the code span $\mathbf{\delta}$ corresponding to $range(n)$ and replacing substrings corresponding to the spans of all child nodes in $C$ with the special token `\str{\#}'.

As illustrated in Figure~\ref{fig:sast}, given the three AST transformation functions \textit{addRange}, \textit{collapseNodes}, and \textit{labelNodes}, if $p$ is a program with AST $A$, the simplified abstract-syntax tree of $p$ is then:
\begin{align*}
    \sast{}(p) = {\rm\it labelNodes}({\rm\it collapseNodes}({\rm\it addRange}(A)))
\end{align*}

Next, we define some properties of a \sast{}. The size of a \sast{} $S$, defined as $size(S)$, is the total number of nodes in $S$. Number of holes in a \sast{}, denoted by
${\rm\it holes}(S)$ is the total count of incomplete subtrees in \sast{} $S$. It is computed as follows:
\begin{align*}
    h(n) =&~ \lvert \mbox{`\str{\#}' tokens in the \textit{label} of node $n$} \rvert \\
    c(n) =&~ \lvert \mbox{number of outgoing edges for node $n$} \rvert \\
    holes(S) =& \sum_{n \in S} h(n) - c(n)
\end{align*}

\sast{} directly maps nodes to code tokens, facilitating the retrieval of the original program through \textit{in-order traversal}.
While the above definitions describe \sast{} as a transformed AST, it is worth noting that extensions can introduce additional edge types, such as control flow and data flow. As a result, in the following sections, we consider \sast{} as a more general graph representation of code.
\subsection{Problem Statement}
\label{sec:problem-stmt}

Given a set of query API methods, such as $\{$\code{nn.Conv2d},  \code{nn.ReLU}$\}$, \toolX{} aims to find idiomatic code examples from a large corpus of code, demonstrating the simultaneous usage of all the API methods in the query. 
As in prior work~\cite{miningidioms}, we view a \textit{code idiom as a syntactic fragment of code frequently occurring across software projects}. For example,
\begin{codeExampleNoHighlight}
self.feat = nn.Sequential(
    nn.Conv2d(27, 16, kernel_size=3), nn.MaxPool2d(2), nn.ReLU(),
    nn.Conv2d(16, 32, kernel_size=3), nn.MaxPool2d(2), nn.ReLU())
\end{codeExampleNoHighlight}
is a common idiom to build convolutional neural networks demonstrating API methods in the query $\{$\code{nn.Conv2d}, \code{nn.ReLU}$\}$. We define this task more formally below.

Let $Q = \{a_1, \ldots, a_m\}$ be a set of query API methods, and $P = \{p_1, \ldots, p_n\}$ be the corpus of programs from which we want to mine idioms. Let  $G \subseteq G'$ denote that a graph $G$ is a subgraph of a graph $G'$.
A code snippet $p$ is called a \textbf{\emph{code idiom}} if it is a non-trivial code snippet and the \sast{} of $p$, which is a graph, is the subgraph of the \sast{}s of a large subset of programs in $P$.
Formally, $p$ is a code idiom if the set $P_p = \{p_i \mid \sast{}(p) \subseteq \sast{}(p_i) \mbox{ and } p_i \in P\}$ is ``reasonably large'' and $p$ is ``a non-trivial code snippet''.  Note that an idiom is better if the size of $P_p$ is larger, which implies that $p$ frequently appears in programs. Similarly, a code snippet is non-trivial if it has a substantial size. For example, the program \code{pass} could be a frequent program in $P$, but its size is so small that it is a trivial program.  
We discuss how \toolX{} determines the ``reasonably large'' size of $P_p$ and the non-triviality of $p$ in a novel search technique in Section~\ref{sec:neural-search}, a contribution of this paper.
An \textbf{\emph{idiomatic code example}}, given a query $Q = \{a_1, \ldots, a_m\}$ (a set of API methods), is a program $p$ such that $p$ is a code idiom in $P$ and each query API $a_i \in Q$ appears in $p$. Each API method in $Q$ could be treated as a standalone program. Therefore, formally, $p$ is an idiomatic example if \sast{} of each $a_i \in Q$ is a subgraph of the $\sast{}(p)$.
A query $Q$ can have multiple idiomatic code examples. \toolX{}'s goal is to discover a set of such examples given a query of API methods.

\paragraph{\textbf{Relations to Subgraph Matching.}} The problem of mining idiomatic code examples can be reduced to the well-studied subgraph matching problem~\cite{corneil70isomorphism, ullmann1976isomorphism}.
Given two graphs $G$ and $G'$, we say $G$ is \textit{{isomorphic}} to the graph $G'$, that is $G \equiv G'$,  if there exists a bijection $f \colon V_G \rightarrow V_{G'}$ such that $(f(v), f(u)) \in E_G$ if an only if $(v, u) \in E_{G'}$.  In the above definition, we use $V_G$ and $E_G$ to denote the nodes and edges of $G$.
Note that in the case where nodes and edges are labeled (such as labels in the \sast{} representation of programs), the bijection $f$ must also match the labels of nodes and edges. 
The goal of \textbf{\emph{subgraph matching}}, given a query graph $G_q$ and a target graph $G_t$, is to count the number of subgraphs in $G_t$ that are isomorphic to $G_q$. Specifically, we want to compute the cardinality of the set $\{H | H \subseteq G_t \mbox{ and } G_q \equiv H\}$. If the cardinality of the set is 0, then $G_q$ is not a subgraph of $G_t$.

Several scalability challenges to the subgraph matching (or isomorphism) problem are due to its NP-completeness~\cite{cook1971complexity}. Traditional approaches use combinatorial search algorithms~\cite{cordella2004sub, gallagher2006matching} but do not scale to large problem sizes. Existing scalable solutions for subgraph isomorphism~\cite{sun2012efficient} involve expensive pre-processing to store locations of many small 2–4 node components and decomposing the queries into these components. Although this allows scaling to large target graphs, the query size cannot scale to more than a few nodes before query decomposition becomes complex. Therefore, we need a scalable technique to answer subgraph queries fast.

\paragraph{\textbf{Scalable Subgraph Matching.}} Here, we can use Graph Neural Networks (GNNs) to answer such queries approximately and quickly.
Recent advancements in GNNs and neural subgraph matching~\cite{neuromatch} propose techniques to predict whether a query graph is a subgraph of a target graph. \texttt{NeuroMatch} \cite{neuromatch} decomposes graphs into small
overlapping neighborhoods and trains a GNN to learn an ordered embedding that preserves subgraph relations.
They have demonstrated effectiveness in practical tasks such as mining frequent motifs in biology, social networks, and knowledge graphs.

One might wonder how approximation would impact the quality of the mined idiomatic examples. Fortunately, we can tolerate minor mismatches in code idioms because two semantically equivalent code snippets may differ in exact syntax due to differences in variable names and control structures.
In the next section (\ref{sec:neuromatch}), we briefly describe the \texttt{NeuroMatch} approach and focus on how we adapt it for code. 

\subsection{Neural Subgraph Matching for Code}
\label{sec:neuromatch}

\toolX{} builds on \texttt{NeuroMatch}~\cite{neuromatch} to perform approximate subgraph matching for code.
Consider arbitrary query and target programs whose \sast{}s are $G_q$ and $G_t$, respectively. Then, the subgraph matching problem involves predicting if $G_q$ is isomorphic to a subgraph of $G_t$.

\toolX{} works in three stages: (1) a \textit{featurization stage}, where \toolX{} featurizes the graph representations of code for learning, (2) an \textit{embedding stage}, where \toolX{} computes the embedding of each subgraph centered at each node of $G_t$, and (3) a \textit{query stage} where the query graph $G_q$ is compared against the target graph in the embedding space for subgraph isomorphism.

\subsubsection{\textbf{Featurization Stage.}} \sast{} nodes are labeled with strings representing the corresponding code spans. Our learning algorithm uses these labels as primary node features. We convert each label into a real-valued embedding vector by passing it through pre-trained large-language models for code (code-LLM) like \texttt{CodeBert}~\cite{codebert}. Code-LLMs provide general-purpose embeddings useful for downstream NL-PL tasks such as natural language code search and documentation generation.
Using code-LLMs to featurize \sast{} nodes, we capture code semantics in embeddings, supplementing the structural graph features. For instance, nodes labeled \code{fp} and \code{file} will map to similar embeddings as they are interchangeably used to refer to a file pointer in code.
We further enhance each node with structural features like node degree, clustering coefficient, and page rank, known to improve GNNs~\cite{neuromatch}.

In the following sections, we use the notation $G_i^j$ to denote a radial $k$-hop \textit{neighborhood} anchored at node $j$ in graph $G_i$.

\subsubsection{\textbf{Embedding Stage.}}
\label{sec:neuromatch-embed}
First, we decompose the target graph $G_t$ into small overlapping radial $k$-hop  \textit{neighborhoods} $G_t^u$ anchored at a node $u$, for each node $u$ in $G_t$. The GNN then generates an embedding $z_t^u$ for each $G_t^u$.
Given the target graph node embeddings $z_t^u$ and an anchor node $v \in G_q$, the subgraph prediction function determines if $G_t^u$ is a subgraph isomorphic to $G_q^v$; i.e., $v$’s $k$-hop neighborhood in $G_q$. Notably, the prediction solely relies on the node embeddings $z_q^v$ and $z_t^u$ of nodes $v$ and $u$, respectively.
We enable this during training by enforcing the embedding geometry to encode the subgraph relation between graphs and learn \textit{order embeddings}~\cite{mcfee2009partial}.
Order embeddings ensure that the subgraph relations are reflected in the embeddings: \textbf{if $G_q^v$ is
a subgraph of $G_t^u$, then the embedding $z_q^v$ of node $v$ is element-wise less than or equal to $u$’s embedding $z_t^u$}:
\begin{equation}
    z_q^v[i] \leq z_t^u[i] ~~~~~\forall_{i=1}^{D} \iff G_q^v \subseteq G_t^u 
\end{equation}
where $D$ is the embedding dimension. We learn \textit{order embeddings} using a max-margin loss function that encodes the above order constraint. Formally, for arbitrary query and target embeddings $z_q$ and $z_t$, we define the loss as:
\begin{equation}
    \mathcal{L}(z_q, z_t) = \sum_{(z_q, z_t) \in P} E(z_q, z_t) + \sum_{(z_q, z_t) \in N} \max\{0, \alpha - E(z_q, z_t)\}
\end{equation}
\begin{equation}
    \text{where}~ E(z_q, z_t) = \norm{\max\{0, z_q - z_t\}}_2^2
\end{equation}

Here, $P$ denotes the set of positive examples where $z_q$'s corresponding neighborhood is a subgraph of $z_t$'s, and $N$ denotes the set of negative examples. For positive examples, the loss is proportional to $E(z_q, z_t)$, which denotes the magnitude of the violation of the order constraint. For negative examples, the amount of violation $E$ should be at least $\alpha$ to have zero loss.

\paragraph{\textbf{Training Data}} To achieve high generalization, we train the GNN with randomly generated query and target graphs.
Let $G_t$ be a graph in our training set.
Positive examples $P$ are constructed by randomly sampling a target neighborhood $G_t^u \subseteq G_t$ anchored at node $u$, and a query $G_t^v \subseteq G_t^u$ anchored at node $v$ using random breadth-first (BFS) walks. When sampling $G_t^v$, we walk $G_t^u$ starting at $u$, ensuring the existence of a subgraph isomorphism mapping from $v$ to $u$. Negative examples $N$ are created by randomly selecting \textit{different} nodes $u$ and $v$ in $G_t$ and performing random traversals.
We use a $k$-layer GNN to embed node $u$, which captures the $k$-hop neighborhood $G_t^u$.
The choice of the number of layers, $k$, depends on the query graph size. For the idiom search task, a $k$ in the range $[7, 10]$ works well. 

Overall, the GNN's message passing ensures that embedding node $u$ is equivalent to embedding the neighborhood $G_t^u$ centered at $u$. Consequently, comparing two node embeddings $z_t^u$ and $z_t^v$ essentially compares the structure of subgraphs $G_t^u$ and $G_t^v$. This property is crucial for the query stage, where predictions are made using a simple element-wise less-than comparison between two embeddings at negligible cost.

\subsubsection{\textbf{Query Stage.}} At the query stage, we determine if $G_q$ is a subgraph of $G_t$. 
We construct a prediction function $f(z_q^v, z_t^u)$ that predicts whether the $k$-hop neighborhood $G_q^v$ anchored at node $v$ is a subgraph of the $k$-hop neighborhood $G_t^u$ anchored at node $u$. This prediction implies that node $v$ maps to node $u$ in the subgraph isomorphism mapping.
We define $f(z_q^v, z_t^u)$ as a function of $E(z_q^v, z_t^u)$, representing the magnitude of violation of the order constraint and the subgraph relation.
A small violation (under a threshold $t$) indicates that the query is likely a subgraph of the target, as the embeddings align with the order constraint:
\begin{equation}
    f(z_q^v, z_t^u) = \begin{cases}
        1 & \text{iff}~ E(z_q^v, z_t^u) < t \\
        0 & \text{otherwise}
    \end{cases}
\end{equation}

Note, while \texttt{NeuroMatch}~\cite{neuromatch} uses a fixed user-defined threshold $t$, we update this with a learnable threshold. We do this by training a multi-layer perceptron (MLP) classifier on top of our GNN during training that makes a binary prediction. We train this classifier with a traditional negative log-likelihood loss function.
\subsection{Neural-guided Idiom Search}
\label{sec:neural-search}
 
We describe a novel search technique that uses neural-subgraph matching to solve the idiomatic example search problem described in Section~\ref{sec:problem-stmt}. We start by formalizing some extended properties of \sast{}s (Section~\ref{sec:sast}) and order embeddings (Section~\ref{sec:neuromatch}). 
First, we observe that while neural-subgraph matching predicts if $G_q$ is a subgraph of $G_t$, it also enables estimating the frequency of a $G_q$ in a set of target graphs. We define the frequency of a query graph in a set of graphs as follows:

\subsubsection{\textbf{Estimating Frequency of Graphs}.}

Assume $g$ and $f$ to be the GNN encoder and the classifier of the pre-trained neural-subgraph matching model (Section \ref{sec:neuromatch}), respectively. 
Let $G_q^v$ denote the query graph anchored at node $v$ and $\mathcal{T}$ be a set of target graphs.
Note that the query graph could be disconnected. For instance, a \sast{} with multiple independent API methods. Consequently, the frequency estimate should ensure all connected components of the query are present in the target.

Let $decompose(G)$ be a function that decomposes a graph $G$ into radial $k$-hop \textit{neighborhoods} $G^u$ anchored at node $u$, $\forall u \in G$.
Then, the frequency of the query graph $G_q^v$ in target graphs $\mathcal{T}$ is the number of neighborhoods that contain \textit{all connected components} of the query graph:

\begin{equation}
\label{eq:freq}
{\rm\it freq}(G_q^v, \mathcal{T}) = 
    \sum_{\forall G_t \in \mathcal{T}}~~~ 
        \sum_{\forall G_t^u \in {\rm\it decompose}(G_t)}
            \gamma(G_q^v, G_t^u)
\end{equation}
\begin{align*}
    \text{where}~~~ \gamma(G_q^v, G_t^u) = \mathbb{I}\left(\forall G_c \in {\rm\it CC}(G_q^v), \quad f(g(G_c), g(G_t^u)) = 1\right)
\end{align*}

Here, ${\rm\it CC}$ returns connected components of a graph, and $\mathbb{I}$ is an indicator function that returns $1$ if all elements in the argument set are $1$, $0$ otherwise.
Note that frequency is estimated here at the neighborhood (subgraphs anchored at some node) level, proportional to the frequency at the graph level.
To efficiently compute $freq$ for a large set of target graphs $\mathcal{T}$, we implement it using the following optimizations:

\begin{enumerate}
    \item \emph{Offline Pre-processed Embeddings}: First, we decompose all graphs $G_t \in \mathcal{T}$ (our search space) into radial k-hop neighborhoods $G_t^u$ anchored at each node $u \in G_t$. Additionally, we can randomly sample nodes $u$ from $G_t$ to compute the neighborhoods. Then, each neighborhood is embedded using the GNN encoder $g$ and stored offline.
    \item \emph{Online Batched Inference}: Second, during inference, the embedding for the
    query graph $g(G_q^v)$ is computed and compared against all pre-processed target embeddings using traditional batched processing on a GPU—making it efficient and scalable.
\end{enumerate}

Next, we observe that \sast{}s for programs can be grown idiomatically by adding nodes to the graph 
that optimize a user-defined metric for the ``idiomaticity'' of a program.

\subsubsection{\textbf{Growing \sast{}s Idiomatically}.}\label{sec:sast-grow}

Suppose we want to search for idiomatic code examples for an API method $a$ in a corpus of programs $P$. The search can be reformulated as generating a program starting from $a$ by ``idiomatically'' adding tokens to the program. A token can be idiomatically added to a program if the token makes the resulting program more frequent in $P$ than by adding any other token. If a program is represented as a sequence of tokens, then such additions would be restricted to the end of the token sequence. However, if we use the \sast{} representation of a program, we can equivalently add a node to the \sast{} ``idiomatically''. The advantage of doing so is that a node can be added at any location in the program. We describe how \toolX{} grows a \sast{} by idiomatically adding nodes to a program:

Let $p$ be an arbitrary program with $\sast{}(p) = S$.
Let $S_k$ be a random subgraph of $S$ with ${\rm \it size}(S_k) = k$ and ${\rm \it frontier}(S_k)$ be the set of nodes $u \in S \setminus S_k$ with an edge to nodes $v \in S_k$.
Let $\mu(S) = \nicefrac{{\rm\it freq}(S, \mathcal{T})}{{\rm\it holes}(S)}$
be a metric of ``\textit{idiomaticity}'' that \toolX{} maximizes while growing \sast{}s.
We use this metric as maximizing $\mu$ involves maximizing the frequency while minimizing 
the holes (incomplete subtrees) in the resulting \sast{}. As a result, \toolX{} optimizes
for the \textit{reusability} and \textit{completeness} of programs.
Then, \toolX{} grows $S_k$ idiomatically by adding a node from its \textit{frontier} that maximizes $\mu$:
\begin{align}
    S_{k+1} \leftarrow S_k \cup u^* \mid u^* = \argmax_{u \in {\rm\it frontier}(S_k)} \mu(S_{k+1})
\end{align}

Note that the node added is in the frontier of $S_k$, which means it has an edge to some node in $S_k$. Therefore, the operation of including a node adds the corresponding edge as well.


\subsubsection{\textbf{Search Algorithm.}}
\toolX{} utilizes the insights above in a search algorithm for idiomatic code examples.
The search starts with some skeleton graphs, called \emph{seed} graphs, where each seed graph is from the program corpus and contains all query API graphs. We cannot use the set of query graphs as a seed graph because such a graph will be disconnected with no frontier and cannot be directly expanded into code examples. The seed graphs are their equivalents, enabling \toolX{} to grow and map them to idiomatic code examples. We next describe the algorithm in detail.

Let $Q = \{a_1, \ldots, a_m\}$ be a set of query API methods and $P = \{p_1, \ldots, p_n\}$ be a corpus of programs from which we want to mine code examples.
Let $\mathcal{Q}$ and $\mathcal{T}$ be the set of \sast{}s corresponding to $Q$ and $P$, respectively.
%
%
\toolX{} begins the search by finding seed graphs, denoted as $\mathcal{S}$, drawn from the set $\mathcal{T}$.
Each seed graph is a candidate graph that contains all query API graphs in $\mathcal{Q}$.
Since API methods (and their \sast{}s) are small, traditional search methods such as regex, text-based search, or combinatorial methods for subgraph matching suffice. In our implementation, \toolX{} uses a efficient 2-phase approach to construct $\mathcal{S}$:
\begin{enumerate}
    \item ElasticSearch\footnote{\href{https://github.com/elastic/elasticsearch}{https://github.com/elastic/elasticsearch}} is an open-source full-text search engine based on Apache Lucene~\cite{lucene}. \toolX{} uses ElasticSearch to build an offline index of the large program corpus $P$. \toolX{} then queries the index for a set of candidate programs $P_{e} \subseteq P$ that contain all API methods in $Q$.
    \item
    \toolX{} then finds the graphs $\mathcal{T}_{e} \subseteq \mathcal{T}$ that correspond to the candidate programs $P_{e}$.
    For each candidate graph $G_t \in \mathcal{T}_{e}$, \toolX{} uses combinatorial subgraph matching~\cite{cordella2004sub} to locate a subgraph in $G_t$ that contains all query API graphs $G_q$ in $\mathcal{Q}$. These subgraphs are the set of seed graphs $\mathcal{S}$.
\end{enumerate}


\toolX{}'s search then grows the seed graphs $G_s \in \mathcal{S}$ towards idiomatic code examples. As described, \toolX{} adds a node to the seed graph from its \textit{frontier} that maximizes an ``\textit{idiomaticity}'' metric $\mu(G_s) = \nicefrac{{\rm\it freq}(G_s, \mathcal{T})}{{\rm\it holes}(G_s)}$.
Iteratively performing this growing procedure an arbitrary $k$ number of times for each seed graph in $\mathcal{S}$ generates graphs (\sast{}s) of size $size(S)+k$. \toolX{} maps these graphs to idiomatic code examples using an in-order traversal.

\subsubsection{\textbf{Search Optimizations}} While simple and complete, this search algorithm can be inefficient if implemented na\"ively. 
To make it efficient and scalable, \toolX{} employs a \textit{beam search} strategy to prune the search space and avoid growing all seed graphs.
Then, \toolX{} identifies a unique \textit{convergence condition} that decides when to stop growing seed graphs.

\paragraph{\textbf{Beam Search}}
Beam search, a heuristic algorithm \cite{beamsearch}, selectively expands the most promising nodes in graph exploration. It is a version of best-first search, organizing partial solutions based on a heuristic. \toolX{} employs beam search to limit the search space by retaining the top $b$ seed graphs with the highest $\mu$ score after each iteration, where $b$ represents the beam width.
\begin{wrapfigure}{r}{0.5\textwidth}
    \centering
    \includegraphics[]{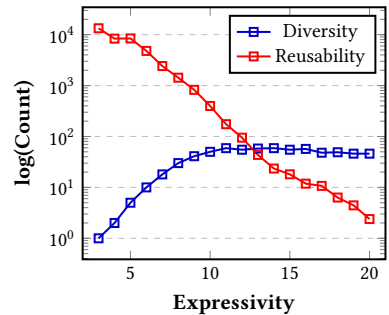}
    \vspace{-1em}
    \caption{\toolX{} tracking \textit{Diversity} and \textit{Reusability} of generated usage examples and reaching the convergence condition for \code{np.mean} at \textit{Expressivity} (size) 12.}
    \label{fig:search-equilibrium}
    \vspace{-1em}
\end{wrapfigure}

\paragraph{\textbf{Convergence Condition}}

While beam search helps prune the search space, we must decide when to stop growing graphs.
To identify a stopping criteria, we observe some properties of an ideal set of idiomatic code examples:

\paragraph{(1) Reusability:} Idiomaticity is inherently related to the notion of \textit{reusability}. An ideal set of usage examples should be \textit{reusable} as building blocks in many scenarios. Our approach provides a natural way to estimate reusability as the average frequency of the \sast{}s returned after each iteration of the search algorithm.

\paragraph{(2) Expressivity:} Secondly, we want usage examples that are \textit{expressive} enough 
to show a complete usage of the APIs of interest. The $size$ of the \sast{}s returned after each iteration of the search algorithm is a good proxy for this.

\paragraph{(3) Diversity:} Lastly, an ideal set of examples should be diverse.\@
Usage examples exhibit clusters, such as standalone usages like \code{np.mean(x, axis=0)},
and co-usage with other APIs like \code{round(np.mean(t), 5)}.
We measure \textit{diversity} by the number of distinct clusters of usage examples returned. Our implementation uses Weisfeiler Lehman (WL) graph hash~\cite{wlhash} to find clusters in \sast{}s returned after each iteration. The hash function iteratively aggregates and hashes the neighborhoods of each node. WL hashes are identical for isomorphic graphs and guarantee that non-isomorphic graphs will get different hashes.

\toolX{} measures these properties during the search and identifies a convergence condition.
The search converges at an ideal \textit{expressivity} (size) when the \textit{reusability} (frequency) and \textit{diversity} (number of distinct clusters) of the usage examples returned after each iteration reach an \textit{equilibrium}.
Figure~\ref{fig:search-equilibrium} illustrates this for \code{np.mean}. Note that our goal is to maximize expressivity while also keeping reusability and diversity high.
The diversity of usage examples increases with the size of the returned \sast{}s because larger 
\sast{}s have more nodes to explore, leading to greater diversity.
Conversely, reusability is inversely proportional to the size of the returned \sast{}s, as larger \sast{}s are less frequent in the corpus.
Once the reusability and diversity converge, continuing the search any further will result in more distinct clusters (diversity) than their frequencies (reusability). Hence, \toolX{} stops the search at this convergence condition.
\section{IMPLEMENTATION}\label{sec:implementation}

We implement our approach in a tool called \toolX{} in \texttt{Python} and 
release it as open-source.\footnote{\href{https://github.com/tart-proj/codescholar}{https://github.com/tart-proj/codescholar}}

\paragraph{\textbf{Data}} To build \toolX{}, we first mine a large corpus of \texttt{Python} code snippets from public GitHub repositories
where \toolX{} can search for idiomatic usage examples.
In our mining process, we exclude forked repositories, repositories with less than ten stars, 
and repositories with the last commit date before 2020. That leaves us with $\approx$85K repositories that amount
to $\approx$4M python files.
We then filter on a set of top-$6$ most popular libraries used in the mined repositories: 
\texttt{pandas}, \texttt{numpy}, \texttt{os}, \texttt{sklearn}, \texttt{matplotlib}, and \texttt{torch}.
Next, we filter out files larger than $1$MB in size or have $<10$ tokens.
This filtering leaves us with $\approx$1.5M python files.
Finally, we extract all the methods/functions from the filtered files and filter out methods/functions
that contain the top-$100$ most popular APIs from each of the top-$6$ libraries. We find these APIs by searching
for unigrams matching API calls and sorting them by frequency. This procedure results in a large
corpus of $\approx$1.4M methods/functions that \toolX{} searches.

\paragraph{\textbf{Configs}} While our training procedure and model architecture is described in Section~\ref{sec:neuromatch-embed},
here, we point out some additional details.
We train our graph neural network using the Adam optimizer with a learning rate of $1 \times 10^{-4}$. We train our model for $5$ iterations on a single NVIDIA GeForce RTX 3080 Ti GPU.\@
Each iteration involves training on the entire dataset in a batched fashion. We use a batch size of $64$. 
%
We use a beam width of $10$ for our search algorithm and employ the 
proposed convergence condition to stop the search automatically.
\section{EVALUATION}\label{sec:evaluation}

This section aims to assess how good \toolX{} generated code examples are.
To do so, in Section~\ref{sec:user-study}, we evaluate if developers prefer \toolX{} examples over
results from state-of-the-art language models like \texttt{GPT3.5}.
In Section~\ref{sec:eval-search}, we then analyse \toolX{} results and quantify how closely
the generated examples represent real-world usage.
Then, in Section~\ref{sec:eval-neuromatch}, we perform some ablations
showcasing \toolX{}'s contributions to neural-subgraph matching for code.
Lastly, in Section~\ref{sec:eval-rag}, we show that \toolX{} not only helps developers, 
but also AI programming assistants in a retrieval-augmented program synthesis setting.

\subsection{Survey with Developers}\label{sec:user-study}

We conducted a survey to measure the quality of code examples generated by \toolX{},
compared to state-of-the-art code generation language models such as \texttt{GPT3.5}~\cite{gpt3, gpt4}.
We follow a similar survey structure as in prior work such as~\cite{egpaper}.
We invited $30$ developers working in research labs and software companies to be interviewed for the survey. With a response rate of $70\%$, $21$ developers completed the survey. We conducted $\approx$10-minute semi-structured interviews with these participants.

\begin{figure*}[t]
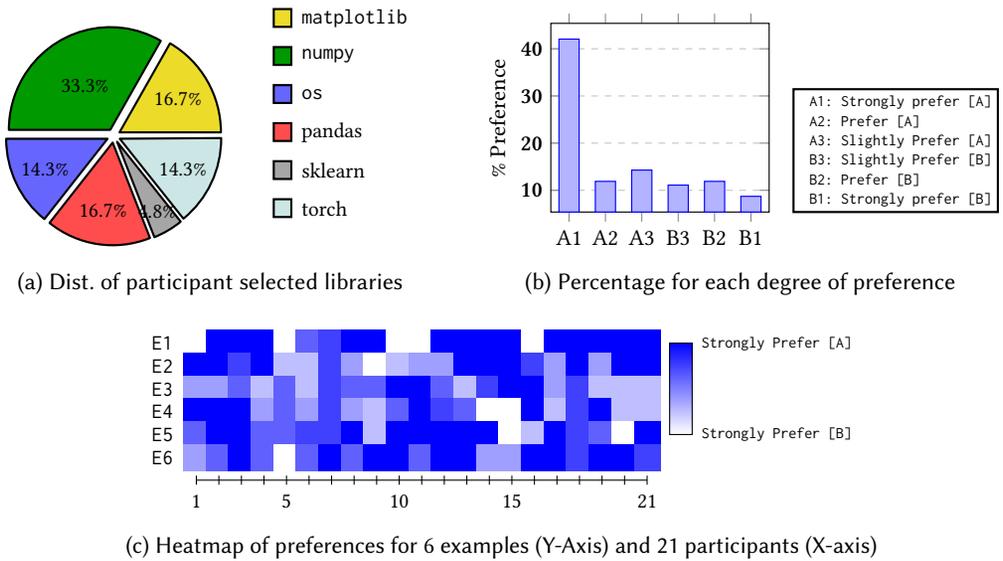

    \centering
    
    \subfloat[Dist. of participant selected libraries]{
	{
            \includegraphics{Figures/user-study-libs}
	}
        \label{fig:study-lib-dist}
    }
    \hspace{1em}
    \subfloat[\centering Percentage for each degree of preference]{
        {
            \includegraphics{Figures/user-study-responses}
        }
        \label{fig:study-pref-dist}
    }\\
    \vspace{1em}
    \subfloat[Heatmap of preferences for 6 examples (Y-Axis) and 21 participants (X-axis)]{
        {
            \centering
            \includegraphics{Figures/user-study-matrix}
        }
        \label{fig:study-pref-matrix}
    }
    \caption{Results of the survey on preferences between \toolX{} and \texttt{GPT3.5}. (a) shows the distribution of libraries selected by the survey participants. (b) shows the breakdown of the degree of preferences chosen by participants (higher is better). (c) shows a heatmap of the individual responses of $21$ participants for each of the $6$ examples they were asked to evaluate. In (c), darker cells are better for \toolX{}. In (b) and (c) \texttt{[A]} and \texttt{[B]} represent \toolX{} and \texttt{GPT3.5}, respectively.}
    \label{fig:study-results}
\end{figure*}


The survey displayed six popular Python libraries and asked participants to select two they were most familiar with.
Figure~\ref{fig:study-lib-dist} shows the distribution of the libraries chosen by participants.
The survey then showed three API methods in the two selected libraries. For each, two code examples were listed: a randomly selected example generated by \toolX{} (Option A) and a randomly selected example generated by \texttt{GPT3.5} (Option B). We use random sampling to avoid any selection biases.
Participants were asked three questions about these examples, which we describe next.

\Paragraph{Q1. Suppose you were learning to use this library. Which example would you prefer to understand the API's complete usage?} Participants were asked to respond with either Option A or B and a degree of preference: Strongly prefer, Prefer, and Slightly Prefer.
%
%
Figure~\ref{fig:study-pref-dist} shows the distribution of participant preferences.
Overall, in $68.3\%$ of the cases, \toolX{} was preferred (either of the 3 degrees of preference) over \texttt{GPT3.5}. Notably, a significant $42.1\%$ of the time \toolX{} examples were strongly preferred over \texttt{GPT3.5}. On the other hand, only $8.73\%$ of the time \texttt{GPT3.5} examples were strongly preferred. 
Figure~\ref{fig:study-pref-matrix} shows a heatmap of the individual preferences of all $21$ participants over
the $6$ examples (3 for each library chosen) they viewed.
Notably, the darker the cells, the stronger the indication that participants preferred \toolX{}'s examples.
It is thus evident from the heatmap that participants consistently favored \toolX{} over \texttt{GPT3.5}.
In fact, $15$ participants preferred more \toolX{} examples than \texttt{GTP3.5}'s, and $6$ had an equal preference for both; i.e., none favored a majority of \texttt{GPT3.5} examples.
Participants described while choosing that \toolX{} examples were clear, grounded, and covered more meaningful behaviors of the API.

\begin{quote}
    $P_{15}$: \textit{``[$A$] example for \code{\small{os.mkdir}} is so cool! It explains not just what the API does but also what it does not (fails when path exists)''}
\end{quote}

\begin{quote}
    $P_{17}$: \textit{``Both [$A$] and [$B$] have information beyond the API. But [$A$] seems to have the right 
    context, whereas [$B$] is vague or irrelevant and could be confusing''}
\end{quote}

\Paragraph{Q2. Which category of examples is more representative of real-world usage?}
Notably, $100\%$ of the participants said that \toolX{} examples (Option A) were more realistic and representative of real-world scenarios. Some independently remarked that the \textit{realism} of \toolX{} generated examples made them much more useful for understanding the API:

\begin{quote}
    $P_5$: \textit{``[$A$] feels more realistic, which allows me to abstract away the key concepts of the API, whereas [$B$] has hardcoded inputs that are just difficult to process.''} 
\end{quote}

\begin{quote}
    $P_8$: \textit{``[$A$] having real and familiar variables and functions around the API's usage makes it more meaningful.''} 
\end{quote}

\Paragraph{Q3. What do you like or dislike about the examples shown?}
Most participants reiterated that they like the realism of \toolX{} generated examples. Providing a familiar or realistic usage context over simple API inputs helps them understand the API better. Some even remarked that having random inputs in examples is noisy and adds additional cognitive load:

\begin{quote}
    $P_4$: \textit{``I actually don't need the dummy dataframe that [$B$] has; it is a distraction because my data will be different anyway.''} 
\end{quote}

Some disliked that unlike \toolX{}, \texttt{GPT3.5} examples, while simple to read and execute, focused a lot on input and output values without any context:

\begin{quote}
    $P_{21}$: \textit{``Example [$B$] for \code{\small{np.dot}} having two dummy \texttt{numpy} arrays as inputs tells me very little about the API. A lot of \texttt{numpy} methods fit the same inputs.''}
\end{quote}

\begin{quote}
    $P_{18}$: \textit{``For \code{\small{np.dot}}, [$B$] is definitely more runnable. But, the \code{\small{np.dot(W, x) + b}} in [$A$] is such a familiar expression that I understand what the API does instantly.''}
\end{quote}

Participants also pointed out that \toolX{} examples frequently showed integration or inter-usage with other APIs, which added to their usefulness:

\begin{quote}
    $P_7$: \textit{``I like that I can search for \code{\small{plt.ylim}}, but get examples where it is used with \code{\small{plt.xlim}}---much like how I would plot data in the real world!''} 
\end{quote}

Lastly, a few participants found that \texttt{GPT3.5} generated examples, while too simple to explain the API's behavior statically, could be executed more easily for further inspection.

\subsection{Analysis of Generated Examples}\label{sec:eval-search}

In Section~\ref{sec:user-study}, our user study highlights various favorable qualities of code examples generated by \toolX{}.
Many participants were particularly interested in the \textit{realism} of the examples, reflecting usage patterns in real-world scenarios.

Here, we propose a novel and meaningful metric to quantitatively evaluate the realism or representativeness of tool-generated code examples.
We capture the notion of realism as the distance between the distribution of \textit{real-world} usage examples and the distribution of
\textit{tool-generated} usage examples.
We can measure such a distance using the \textit{Earth Mover's Distance} (EMD)~\cite{rubner2000earth} metric.
Intuitively, EMD measures the minimum cost of transforming one distribution into another, and a lower EMD implies
the distributions are closer.
Intuitively, this measures \textit{``How close are the tool-generated examples to real-world scenarios?''}


\paragraph{\textbf{Setup}}

We first define the two distributions to measure the EMD between the distribution of \textit{tool-generated} and \textit{real-world} usage examples. For the \textit{real-world} distribution, we pick \textit{all} code snippets
in our large mined corpus (Section~\ref{sec:implementation}) that contain the query APIs.
We then convert these snippets into vectors using OpenAI's GPT-3.5~\cite{gpt3} Embedding API 
\footnote{\str{text}-\str{embedding}-\str{ada}-\str{002}} and aggregate them as a uniform distribution.
Similarly, we embed \textit{tool-generated} examples into a distribution of vectors. 
Finally, we compute the EMD between the two distributions.

We evaluate EMD in multiple usage settings of \toolX{}.
First, we evaluate single API queries. This models a setting where developers initially explore
and learn individual APIs before using them in conjunction with others. It also
asses how \toolX{} generated examples can augment existing API documentation. Here, we evaluate on
a set of $60$ API methods from $6$ popular python libraries: 
\texttt{pandas}, \texttt{numpy}, \texttt{os}, \texttt{sklearn}, \texttt{matplotlib}, and \texttt{torch}.
We select the top-$10$ API methods for each library by inspecting unigram frequency in the mined corpus.

Second, we evaluate multi-API queries, a common scenario where developers
use multiple APIs in conjunction with code for orchestration or integration. We evaluate on
a set of $25$ multi-API queries, categorized as follows.
\texttt{\small{single-library-pairs}} are $15$ queries with API pairs from the same library.
\texttt{\small{mixed-library-pairs}} are $5$ queries with API pairs from different libraries.
Lastly, \texttt{\small{triplets}} are $5$ queries that contain three APIs.
We identify these queries by inspecting frequencies of API bigrams (pairs) and trigrams (triplets) 
for \texttt{pandas}, \texttt{numpy}, \texttt{os}, \texttt{matplotlib}, and \texttt{torch} in our corpus.
We exclude \texttt{sklearn} because its APIs were mostly used in isolation. We consider the top-$3$ bigrams per library for \texttt{\small{single-library-pairs}}, and an overall top-$5$ for \texttt{\small{mixed-library-pairs}} and \texttt{\small{triplets}}.

\paragraph{\textbf{Baselines}}
We compare \toolX{} with two baselines: \texttt{Rand} and \texttt{GPT3.5}.
In \texttt{Rand}, we randomly select examples from the \textit{real-world} distribution.
By comparing against \texttt{Rand}, we aim to show that EMD captures the non-triviality
of generating idiomatic and realistic code examples, particularly that \toolX{} goes beyond simple random sampling 
of code snippets from the real world. 
Second, we compare with \texttt{GPT3.5} (\str{gpt}-\str{3.5}-\str{turbo}), a state-of-the-art large-language model (LLM) 
for code generation tasks. LLMs like \texttt{GPT3.5} excel in producing idiomatic (commonly used) code.
Note, we compare against \texttt{GPT3.5} over \texttt{GPT4} because it is cheaper and powers the popular user-facing tool ChatGPT.
We carefully engineered prompts for \texttt{GPT3.5} to return real-world idiomatic examples given the APIs of interest. 
This prompt engineering process, which \toolX{} doesn't require, is essential due to LLMs' sensitivity to formatting and instructions.
For \texttt{Rand} we choose as many examples as \toolX{} generates for a query, and the \texttt{GPT3.5} prompt
imposes no constraints on the number of examples to generate, ensuring a fair comparison.
Note that we don't compare with E.G.~\cite{egpaper} because it's closed-source, even though it supports single API queries.

\begin{figure}
    \centering
    \includegraphics[]{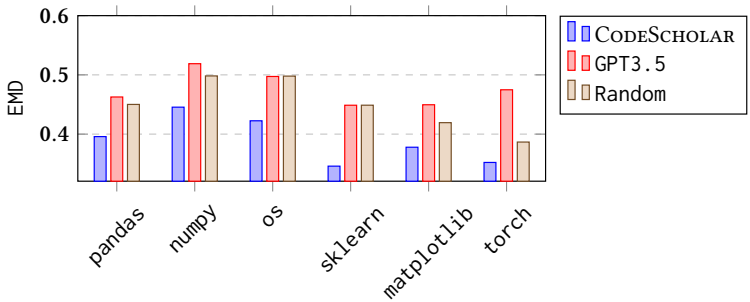}
    \vspace{-1em}
    \caption{Average Earth Mover's Distance (EMD) between \textit{tool generated} API usage examples and \textit{real-world} usage examples for single API queries for various libraries. Lower EMD is better, implying the generated examples are closer to real-world usage. Y-axis: EMD, X-axis: Python Library}
    \label{fig:single-api-emd}
\end{figure}

\subsubsection{\textbf{Single API Queries}}\label{sec:eval-single}

Figure~\ref{fig:single-api-emd} shows the average EMD for APIs from the $6$ libraries. 
Note again that a distribution with a lower EMD implies that the examples generated are more representative of the actual usage than one with a higher EMD. 
Overall, we observe that \toolX{} has a low average EMD of $0.39$ compared to a $0.47$ for \texttt{GPT3.5} and $0.45$ for \texttt{Rand}.
\toolX{} examples have a lower EMD than both \texttt{GPT3.5} and \texttt{Rand} across all libraries, showcasing that they are more realistic.
Notably, a higher EMD for \texttt{Rand} than \toolX{} indicates that generating diverse and representative usage examples is more non-trivial than randomly picking some examples from \textit{real-world}. 
%
%
Below, we discuss a few usage examples generated by \toolX{} vs \texttt{GPT3.5}.

\begin{example} Suppose we have a trained deep-learning classifier that returns a tensor of predicted probabilities for each class during inference. Here, we would use the \code{torch.argmax} API to get the class with the highest probability. \texttt{GPT3.5} returns the following example for \code{torch.argmax}:

\vspace{-\abovedisplayskip}
\begin{gptExample}
                                            |\textbf{GPT}|
1 tensor1 = torch.tensor([1, 5, 3, 9, 2])
  max_index1 = torch.argmax(tensor1)
\end{gptExample}

On the other hand, here are 2 \toolX{} examples that are observably more \textit{\textbf{representative}} of the usage scenarios for \texttt{torch.argmax}; in fact, they are directly applicable to our task.

\vspace{-\abovedisplayskip}
\begin{codeExample}
                                        |\textbf{\toolX{}}|
1 torch.argmax(logits, dim=(logits.dim() - 1), keepdim=True)
2 preds = torch.argmax(y_hat.detach(), dim=(- 1))
\end{codeExample}
\end{example}
\vspace{-\belowdisplayskip}

\begin{example}Suppose we have a directory containing many text files and want to process each. The processing function \code{foo} takes a file path as input.
Here, we would use the \code{os.listdir} API to get the list of files in the directory and run \code{foo} on it. \texttt{GPT3.5} returns the following $2$ examples:

\vspace{-\abovedisplayskip}
\begin{gptExample}
                                            |\textbf{GPT}|
1 l = [file for file in os.listdir('.') if os.path.isdir(file)]
2 file_list = os.listdir('/path/to/directory')
\end{gptExample}
\vspace{-\belowdisplayskip}

While correct, it does not showcase the API's \textit{\textbf{complete behavior}}. However, one of \toolX{}'s examples captures that \texttt{os.listdir} returns a relative path---more useful for our task.

\vspace{-\abovedisplayskip}
\begin{codeExample}
                                        |\textbf{\toolX{}}|
1 for f in os.listdir('./training_dataset/orange'):
      color_hist_of_image(('./training_dataset/orange/' + f))
\end{codeExample}
\vspace{-\belowdisplayskip}
\end{example}

\vspace{-1.5em}
\begin{example} We also observe that compared to \texttt{GPT3.5}, \toolX{} generates a very \textbf{\textit{diverse}}
set of examples for each API. For instance, here are 3 examples for \code{df.groupby} in \texttt{pandas}:

\vspace{-\abovedisplayskip}
\begin{codeExample}
                                        |\textbf{\toolX{}}|
1 df.groupby('Period').Choice.value_counts(normalize=True).unstack
2 df.groupby(['quarter']).apply(do_one_merger_breakup)
3 df.groupby(['id']).aggregate(take_first_annotation)
\end{codeExample}

Similarly, here are 3 examples for \code{np.matmul} in \texttt{numpy} showing various popular use cases such as computing 
squared-euclidean distance, forward pass of neural nets and pseudo-inverses.

\vspace{-\abovedisplayskip}
\begin{codeExample}
                                        |\textbf{\toolX{}}|
1 np.matmul((X_mean - Y_mean), (X_mean - Y_mean).T)
2 np.matmul(self.x, self.w1) + self.b1
3 np.matmul(np.matmul(np.transpose(X), V_inv), X)
\end{codeExample}
\vspace{-\belowdisplayskip}
\end{example}


\begin{figure*}[!t]
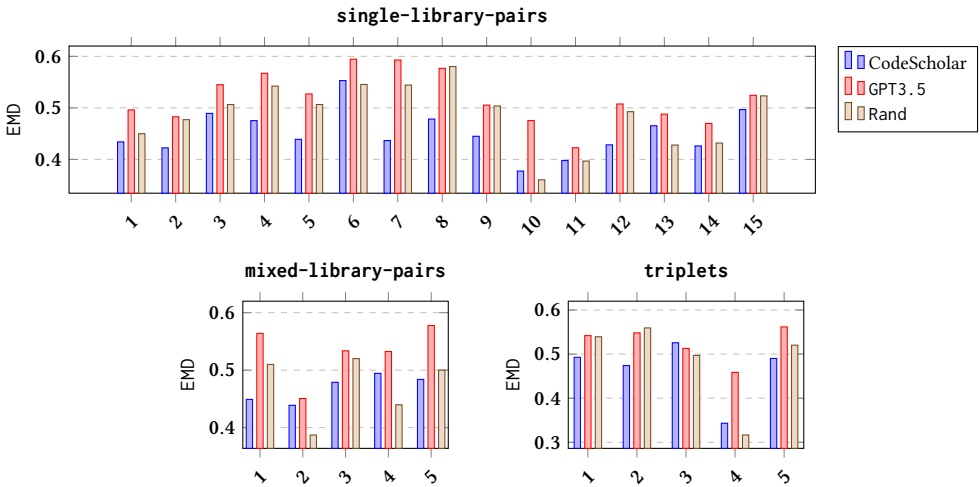

  \captionsetup[subfigure]{labelformat=empty}
  \subfloat[]{\includegraphics[]{Figures/emd-pairs}}\\
  \vspace{-1.5em}
  \subfloat[]{\includegraphics[]{Figures/emd-mixpairs}}\hspace{1.5em}
  \subfloat[]{\includegraphics[]{Figures/emd-triplets}}
  \vspace{-2.5em}
  \caption{Earth Mover's Distance (EMD) between \textit{tool generated} API usage examples and \textit{real-world} usage examples for multi-API queries. \texttt{\footnotesize{single}} and \texttt{\footnotesize{mixed}} \texttt{\footnotesize{library-pairs}} are queries with a pair of APIs from single and multiple libraries, respectively. \texttt{\footnotesize{triplets}} are queries with three APIs. Lower EMD is better, implying the generated examples are closer to real-world usage. Y-axis: EMD, X-axis: API ID}\label{fig:multi-api-emd}
  \vspace{-1em}
\end{figure*}


\subsubsection{\textbf{Multi API Queries}}\label{sec:eval-multi}

Figure~\ref{fig:multi-api-emd} shows the EMD for the three categories of multi-API queries. Overall, similar to Section~\ref{sec:eval-single}, we observe that on average \toolX{} has a lower average EMD $0.45$ compared to a $0.52$ for \texttt{GPT3.5} and $0.48$ for \texttt{Rand}. For $96\%$ (all but 1) of the queries, \toolX{} has lower EMD compared to \texttt{GPT3.5}, implying \toolX{}'s usage examples are more representative for multi-API queries too. We discuss some interesting examples below. 

\begin{example} \code{(np.mean, np.sqrt)} is a query from \texttt{single-library-pairs}, for which \texttt{GPT3.5} returns a good usage example, with details about the input:

\vspace{-\abovedisplayskip}
\begin{gptExample}
                                            |\textbf{GPT}|
1 array = np.array([[1, 2, 3], [4, 5, 6]])
  mean_axis_0 = np.mean(array, axis=0)
  sqrt_mean_axis_0 = np.sqrt(mean_axis_0)
\end{gptExample}
\vspace{-\belowdisplayskip}

Whereas \toolX{} returns a more representative, concrete, and familiar real-world example related to calculating root-mean-squared error:

\vspace{-\abovedisplayskip}
\begin{codeExample}
                                        |\textbf{\toolX{}}|
1 def calc_rmse(true, pred):
    return np.sqrt(np.mean(np.square((true - pred))))
\end{codeExample}
\vspace{-\belowdisplayskip}
\end{example}

\vspace{-1.5em}
\begin{example} Another \texttt{single-library-pair} is \code{(torch.load, torch.save)}. 
\texttt{GPT3.5} returns a long program (>$20$ LOC) that is correct but is too involved.

\vspace{-\abovedisplayskip}
\begin{gptExample}
                                            |\textbf{GPT}|
1 model = MyModel()
  |$\updownarrow$| 4 lines
  for epoch in range(num_epochs):
        |$\updownarrow$| 7 lines
  torch.save(model.state_dict(), 'model.pth')
  loaded_model = MyModel()
  loaded_model.load_state_dict(torch.load('model.pth'))
  |$\updownarrow$| 6 lines
\end{gptExample}

However, \toolX{} finds a \textit{\textbf{concise}} but expressive example:

\vspace{-\abovedisplayskip}
\begin{codeExample}
                                        |\textbf{\toolX{}}|
try:
    ext_tokenizer = torch.load('mbart-large-50.tokenizer.pt')
except FileNotFoundError as e:
    torch.save(ext_tokenizer, 'mbart-large-50.tokenizer.pt')
\end{codeExample}
\end{example}

\begin{example} Finally, we explore a challenging \texttt{triplet} query: \code{(os.path.join, os.makedirs, os.path.exists)}. 
\toolX{} generates examples that demonstrate a prevalent idiom—checking if a directory exists before creating one:

\vspace{-\abovedisplayskip}
\begin{codeExample}
                                        |\textbf{\toolX{}}|
1 def get_config_path(self):
     config_dir = os.path.join(self.parent_dir, self.config_dir_name)
     if (not os.path.exists(config_dir)):
         os.makedirs(config_dir)
\end{codeExample}
\vspace{-\belowdisplayskip}

\texttt{GPT3.5} tends to generate examples that only satisfy a subset of the APIs, go beyond the query APIs, or use non-existing ones, which is referred to as \textit{hallucination}~\cite{gorilla}.

\vspace{-\abovedisplayskip}
\begin{gptExample}
                                            |\textbf{GPT}|
1 project_root = "/path/to/project"
  directories = ["src", "tests", "docs"]
  os.makedirs(project_root)
  for directory in directories:
      os.makedirs(os.path.join(project_root, directory))

2 file_path = "/path/to/file.txt"
  if not os.path.exists(file_path):
      with open(file_path, "w") as file:
          file.write("Hello, world!")
\end{gptExample}
\vspace{-\belowdisplayskip}

\end{example}
\vspace{-1.5em}


\subsection{Evaluating Neural Subgraph Matching for Code}\label{sec:eval-neuromatch}

Next, we evaluate the performance of neural subgraph matching to predict if a query \sast{} 
is a subgraph of a target \sast{}.
While a comprehensive evaluation of general neural subgraph matching is presented in Luo et al.~\cite{neuromatch}, 
we primarily focus on its performance for programs.
Particularly, we are interested in ablations identifying how various components of our adaptation affect the
performance of the GNNs predictions: (1) the choice of graph representation and (2) the choice of node features.

\paragraph{\textbf{Setup}} 

To evaluate our models, we first sample 10,000 programs from the large corpus of code 
we mined from GitHub (described in Section~\ref{sec:implementation}). 
Then, we perform the random sampling procedure described in the \textit{Training Data} section of Section~\ref{sec:neuromatch}.
To summarize, we construct synthetic positive and negative examples for subgraph prediction performing random BFS walks on the \sast{}s of the sampled programs. We then split the resulting data 
into 640000 training (80\%) and 160000 test (20\%) examples.
In Table~\ref{tab:neuromatch-eval}, we report the accuracy (\% of correct predictions), 
recall (\% positive examples correctly predicted), and 
precision (\% positive predictions that were correct) on the test set.

\begin{wraptable}{r}{0.5\textwidth}
    \vspace{-2mm}
    \caption{Ablations for Neural Subgraph Matching for Code. 
            Acc., Prec., and Rec. are accuracy, precision, and recall, respectively.}
    \vspace{-2mm}
    \label{tab:neuromatch-eval}
    \begin{tabular}[t]{lccccl}
    \toprule
         \textbf{Model} 
         & \textbf{Acc.} & \textbf{Prec.} & \textbf{Rec.} & \\
         
    \midrule
        \str{NM}-\str{AST} & 0.80 & 0.70 & 1.00 \\
        \str{NM}-\sast{}-\str{Abstract} & 0.85 & 0.77 & 0.99 \\
        \str{NM}-\sast{} & 0.90 & 0.82 & 1.00 \\
        \str{NM}-\sast{}-\str{CodeBERT} & \textbf{0.96} & \textbf{0.93} & \textbf{1.00} \\
    \bottomrule
    \end{tabular}
\end{wraptable}

\paragraph{\textbf{Results}}
The \str{NM}-\str{AST} model uses
an \str{AST} representation and \str{AST} node types as One-Hot Encoded node features.
This model achieves high accuracy and recall but a poor precision of $0.7$.
We conjecture this results from the \str{AST} being quite large and containing many unnecessary node types that do not carry much information.
Next, we evaluate the \str{NM}-\sast{}-\str{Abstract} model by replacing the \str{AST} with a \sast{}.
However, we continue to use the \str{AST} node types as node features (hence the \str{Abstract} suffix)
to evaluate the effect of simplifying graph structure. We observe that this leads to a significant
improvement in precision of $7\%$. Then we evaluate the \str{NM}-\sast{} model, which uses the \sast{} graph
and the \sast{} node labels as node features encoded using Byte-Pair Encoding (BPE)~\cite{bpe}.
This leads to a further improvement in precision and accuracy of $5\%$.
While BPE effectively encodes tokens of the source code span, 
it lacks a semantic understanding of what the tokens represent.
Consequently, we evaluate the \str{NM}-\sast{}-\str{CodeBERT} model, which uses \sast{} graphs and embeds
the \sast{} node labels using a pre-trained CodeBERT model~\cite{codebert}. This improves our model performance 
to $96\%$ accuracy and $93\%$ precision.
\subsection{\toolX{} for Retrieval-Augmented Generation}\label{sec:eval-rag}

We showed that \toolX{} is effective at generating more representative
and diverse API usage examples compared to LLMs such as \texttt{GPT-3.5}.
Although LLMs are more general-purpose and can generate code for arbitrary tasks given a natural language description, even state-of-the-art LLMs find reasoning and code-related tasks more challenging than others~\cite{gpt3, gpt4, llama2}.
This is especially true with tasks involving APIs, primarily due to their inability to generate accurate input
arguments and their tendency to hallucinate. Recent work has shown that augmenting LLMs with 
retrievers can improve their performance on such tasks~\cite{gorilla}.
Here, we aim to evaluate \toolX{}'s ability to act as a \textit{retriever} for such retrieval-augmented generation (RAG) models.

\begin{wrapfigure}{r}{0.5\textwidth}
    \centering
    \includegraphics[]{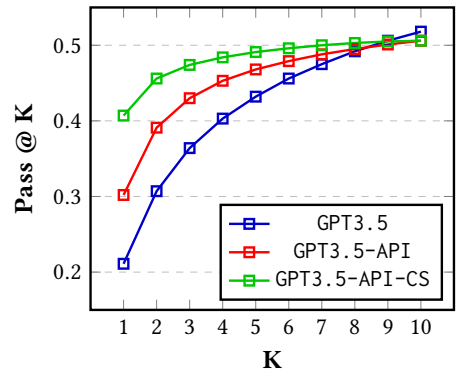}
    \vspace{-1em}
    \caption{Pass@K for the ODEX natural language to code (NL2Code) task (higher is better).
    \texttt{GPT3.5} is a vanilla LLM-based program synthesizer. \texttt{GPT3.5-API} uses chain-of-thought to find relevant APIs before synthesis. \texttt{GPT3.5-API-CS} additionally adds \toolX{} examples for the relevant APIs to the synthesis prompt.}
    \label{fig:eval-rag}
    \vspace{-1em}
\end{wrapfigure}

\paragraph{\textbf{Setup}}
For evaluation, we use ODEX~\cite{odex}—an open-domain execution-based natural language (NL) to code generation dataset.
Note, we do not use other popular benchmarks such as HumanEval~\cite{humaneval}, APPS~\cite{apps}, or MBPP~\cite{mbpp}
because these tasks do not involve the use of APIs.
However, ODEX covers various library APIs and methods in \texttt{Python}.
It contains $439$ examples with NL descriptions, corresponding code snippets, and test cases.
We sample from this $85$ examples that deal with APIs in the top-$6$ libraries mentioned in Section~\ref{sec:implementation}.
We follow Chen et al.~\cite{humaneval} and measure the performance as the execution accuracy using
the $pass@k$ metric—the fraction of problems having at least one correct prediction within $k$ samples.
Correct here means that the generated code, when executed, passes all test cases.

\paragraph{\textbf{Results}}
Figure~\ref{fig:eval-rag} shows the results of our evaluation.
We start by evaluating \texttt{GPT3.5} with no retrieval. Here, we prompt
the model with the NL description and $3$ few-shot examples of the task and ask it to generate the code.
We observe that the model scores quite low ($0.2$)
on $pass@1$, indicating that the model struggles to identify the correct intent. The model, however, expectedly improves to $0.5$ ($pass@10$) when given more tries and variability. 
The next baseline \texttt{GPT3.5-API} first asks \texttt{GPT3.5} to generate
a relevant API method for the given task. Then, the model is asked to generate the code snippet using the generated API method. This prompting style is called \textit{chain-of-thought}~\cite{chainofthought}. 
Adding the API method to the prompt improves $pass@1$ to $0.3$. 
Lastly, we evaluate \texttt{GPT3.5-API-\toolX{}}, which is similar to \texttt{GPT3.5-API} but in addition
to the API method, we also add $10$ \toolX{} generated API usage examples to the prompt; i.e., retrieval-augmented generation.
This improves the performance of \texttt{GPT3.5-API} to $0.4$ ($pass@1$), a $50\%$ relative improvement overall.
Note: we do not intend this to be a comprehensive evaluation of \texttt{GPT3.5} or RAG models. We leave that for future work.
However, our evaluation shows the potential of tools like \toolX{} to improve the performance of LLMs on code-related tasks.

\section{Discussion}

\textit{Extensibility.} \toolX{} searches for idiomatic
usage examples by treating programs as \sast{} graphs. However, the general framework
proposed in this work is independent of the graph representation used. For example,
one could find idiomatic patterns in control-flow graphs or computation graphs
of machine-learning models. Also, \toolX{} is independent of the datasets used in this work.
One can index and search over any dataset (public codebases or private repositories within organizations)
at any granularity using a pre-trained \toolX{}. Further, the \sast{} representation introduced in this work can be constructed for languages other than Python. One could potentially use \sast{} for cross-language code search tasks.

\textit{Generalizing with LLMs.} From the examples in Section \ref{sec:eval-single} and \ref{sec:eval-multi}, 
we observe that while \toolX{} examples are more \textit{representative} of real-world usage, \texttt{GPT3.5} generates
code that is \textit{generic}. Generic code requires minimal edits from the user to be \textit{runnable}.
Future work can explore prompting LLMs to polish \toolX{} generated examples or create inputs to execute them.

\textit{Search vs Generation.} \toolX{} is a search-based approach that grows
program graphs to find accurate idiomatic usage examples limited to examples that occur in the real world.
However, generative approaches (LLMs) can synthesize examples for newer APIs and their interactions with
existing APIs. Also, the generative nature of LLMs allows them to be faster than graph-search-based tools
while sacrificing correctness. Future work can explore developing generative approaches for program graphs.
\section{Related Work}

\paragraph{Code Search.} Prior work has shown that developers often search for code to learn new APIs and 
find code snippets that solve specific tasks~\cite{brandt2009two, montandon2013documenting, sadowski2015developers}.
Developers use several code search tools and engines to find code snippets.
However, most of these tools focus on enriching the query with additional information, 
such as type information~\cite{mandelin2005jungloid} and test cases~\cite{lemos2007codegenie}.
More relevant to our work are code-to-code search engines such as \texttt{FaCoY}~\cite{kim2018facoy}, which take a code snippet as a query
and retrieve relevant code snippets~\cite{kim2018facoy,gu2018deep,wang2010matching}. 
However, unlike our work, these tools are limited to searching code snippets that are similar to the query and are not designed
to help developers learn APIs or find API usage examples.
A recent study on code search~\cite{di2023code} surveyed $30$ years of research, giving a comprehensive overview of challenges and techniques that address them. Interestingly, they find that despite different goals, code search relates to other problems such as code completion~\cite{codex} and clone detection~\cite{roy2007survey}. \toolX{} supports this by reformulating search as a growing/completion task while optimizing properties like frequency of matches.

Efforts to improve keyword-based code search~\cite{sachdev2018retrieval, mcmillan2011exemplar} are adjacent to this work. Portfolio~\cite{portfolio} retrieves functions and visualizes their usage chains. CodeHow~\cite{codehow} augments the query with API calls which are retrieved from documentation to improve search results. CoCabu~\cite{cocabu} augments queries with structural code entities. While these search techniques focus on improving the query, they do not directly support generating idiomatic usage examples for APIs, especially multi-API queries.

\paragraph{Idiom and Pattern Mining.} Tools that mine idioms or idiomatic examples from code repositories are relevant to our work. Allamanis et al.~\cite{miningidioms} propose HAGGIS—a Bayesian approach for mining idioms from code. GraPacc~\cite{grapacc} achieves pattern-oriented code completion by first mining graph-represented coding patterns using GrouMiner~\cite{grouminer}, then searching for input code to produce completions. More recent work \cite{apiretrieval, apibytecode, apicoderec} has improved this by predicting the next API call given a code change.
All these approaches are limited to offline mining of idioms and usage patterns ahead of time and are not interfaces for searching idiomatic code.
In contrast, \toolX{} provides an interface for idiomatic code search and can be extended for offline idiom mining by growing random graphs instead of seed graphs matching the query.
As discussed in Section~\ref{sec:existing-tools} tools like E.G.~\cite{egpaper} are similar to our work. Unlike \toolX{}, E.G. does not support multi-API queries and uses simple stopping criteria, such as constant values for max nodes and min support.
\toolX{}, on the other hand, uses dynamic criteria that identify ideal properties of idiomatic code. Additionally, E.G. is a closed-source tool, which restricted its use in our single-API evaluation.

\paragraph{LLMS for APIs.} More recently, there has been a growing interest in using language models for API-related tasks~\cite{toolformer,huggingpt,taskmatrix,gorilla}.
For example, Patil et al.~\cite{gorilla} finetune a LLaMA~\cite{llama2} model on a large corpus of API documentation of ML APIs. 
These models are limited to specific API domains (e.g., hugging-face APIs), highly dependent on curating good-quality training data, and require significant resources.
\toolX{}, on the other hand, is a generic tool that can search any code corpus, uses auto-generated training data, and is significantly more resource-efficient than LLMs.
Further, as shown in Section~\ref{sec:eval-rag}, \toolX{} can potentially be used to improve the code generated by LLMs.
\section{Conclusion}

We presented \toolX{}, a tool that generates idiomatic code examples demonstrating the common usage of API methods.
\toolX{} is built on the novel insight that treating programs as graphs
reduces idiomatic code search to frequent subgraph matching. Based on this insight,
\toolX{} employs a neural-guided search algorithm over graphs that grows query APIs into idiomatic code snippets.
We showed that developers strongly prefer \toolX{} generated examples over
those from state-of-the-art language models like \texttt{GPT3.5}.
Our quantitative evaluation suggested that, for several single and multi-API queries, \toolX{} generates
realistic, diverse, and concise examples.
Lastly, we demonstrated that \toolX{} can also improve the correctness of LLM-powered assistants
for general program synthesis tasks.

\bibliographystyle{ACM-Reference-Format}
\bibliography{references}

\end{document}